\documentclass[aps,prd,nofootinbib,twocolumn,superscriptaddress]{revtex4}
\usepackage{amsmath, amssymb, amsthm, graphicx, epsfig, fancyhdr,epsfig, slashed, mathrsfs, nccmath}
\makeatletter\usepackage{microtype}\g@addto@macro\@verbatim{\microtypesetup{activate=false}}\makeatother%
\usepackage[T1]{fontenc}
\usepackage{tikzsymbols}
\usepackage{natbib}
\usepackage{float}
\usepackage[thinc]{esdiff}
\usepackage{enumitem}
\usepackage{booktabs,tabularx}
\newcolumntype{Y}{>{\raggedright\arraybackslash}X}

\usepackage{tikz,xcolor,hyperref}
\hypersetup{
    colorlinks=true,
    citecolor=blue,
    linkcolor=brown,
    filecolor=magenta,      
    urlcolor=cyan,
    pdftitle={preheatingGwsDM},
}
\definecolor{lime}{HTML}{A6CE39}
\DeclareRobustCommand{\orcidicon}{
	\begin{tikzpicture}
	\draw[lime, fill=lime] (0,0) 
	circle [radius=0.2] 
	node[white] {{\fontfamily{qag}\selectfont \tiny ID}};
	\draw[white, fill=white] (-0.0625,0.095) 
	circle [radius=0.007];
	\end{tikzpicture}
	\hspace{-2mm}
}

\foreach \x in {A, ..., Z}{\expandafter\xdef\csname orcid\x\endcsname{\noexpand\href{https://orcid.org/\csname orcidauthor\x\endcsname}
			{\noexpand\orcidicon}}
}

\newcommand{\be}{\begin{equation}}
\newcommand{\ee}{\end{equation}}
\newcommand{\bea}{\begin{eqnarray}}
\newcommand{\eea}{\end{eqnarray}}

\def\Mp{M_{\rm Pl}}

\graphicspath{{./}{./Figs/}{./figs/}{../Figs/}{./fig/}}

\begin{document}

\title{Detectable gravitational waves from preheating probes nonthermal dark matter}

\author{Anish Ghoshal}
\email{anish.ghoshal@fuw.edu.pl}
\affiliation{Institute of Theoretical Physics, Faculty of Physics, University of Warsaw,ul.  Pasteura 5, 02-093 Warsaw, Poland}

\author{Pankaj Saha\orcidA{}\footnote{Corresponding author}}
\email{pankaj@post.kek.jp}
\affiliation{Institute of Particle and Nuclear Studies~(IPNS), High Energy Accelerator
Research Organization (KEK), Oho 1-1, Tsukuba 305-0801, Japan}
\affiliation{School of Natural Sciences,
Seoul National University of Science and Technology,
232 Gongneungro, Nowon-gu, Seoul, 01811, South Korea}

\begin{abstract}

\textit{We describe the challenges and pathways when probing inflaton as dark matter with the stochastic gravitational waves (GWs) signal generated during the (p)reheating. Such scenarios are of utmost interest when no other interaction between the visible and dark sectors is present, therefore having no other detectability prospects. 
We consider the remnant energy in the coherently oscillating inflaton's zeroth mode to contribute to the observed relic dark matter density in the Universe. 
To fully capture the nonlinear dynamics and the effects of back-reactions during the oscillation, we resort to full nonlinear lattice simulation with pseudo-spectral methods to eliminate the differencing noises. We investigate for models whose behavior during the reheating era is of $m_{\Phi}^2\Phi^2$ type and find the typical primordial stochastic GWs backgrounds spectrum from scatterings among highly populated inflaton modes behaving like matter. We comment on the challenges of constructing such viable inflationary models such that the inflaton will account for the total dark matter of the universe while the produced GWs are within the future GWs detectors such as BBO, DECIGO, PTA, AION-MAGIS, and CE. We also describe the necessary modifications to the standard perturbative reheating scenario to prevent the depletion of residual inflaton energy via perturbative decay
}
\end{abstract}

\maketitle

\section{\label{intro}Introduction}
The origin and composition of dark matter (DM) in the Universe remains an open problem in modern particle physics and cosmology~\cite{Jungman:1995df,Bertone:2004pz,Feng:2010gw}. Even with great experimental efforts over the last 30 years, no clinching evidence of the existence of thermal frozen out DM dubbed as the famous ``WIMP miracle'' ~\cite{Lee:1977ua,Scherrer:1985zt,Srednicki:1988ce,Gondolo:1990dk} has been found in any DM search experiments ~\cite{PandaX-II:2017hlx,PandaX:2018wtu,XENON:2020kmp,XENON:2018voc,LUX-ZEPLIN:2018poe,DARWIN:2016hyl}, or via indirect detection~\cite{HESS:2016mib,MAGIC:2016xys} or through collider physics searches (e.g.~at the LHC~\cite{ATLAS:2017bfj,CMS:2017zts}). This motivates us to consider alternative DM production, particularly nonthermal production mechanisms, such that the observed DM abundance is generated out of equilibrium by the so-called {\it freeze-in} mechanism~\cite{McDonald:2001vt, Hall:2009bx, Bernal:2017kxu}, or say, via particle production during (p)-reheating~\cite{Moroi:1994rs, Kawasaki:1995cy, Moroi:1999zb, Jeong:2011sg, Ellis:2015jpg,Harigaya:2014waa, Garcia:2018wtq, Harigaya:2019tzu, Garcia:2020eof,Chung:1998zb, Chung:1998ua, Co:2017mop}.

Interestingly, irrespective of the production mechanisms, the expected signals and constraints on the dark matter candidates are dictated by their interactions with the states of the standard model~(SM). In this paper, we consider the scenario in which the DM relic abundance is set via the inflaton oscillations and decay, which is similar to the out-of-equilibrium freeze-in production mechanism. The cosmic inflation resolves the flatness and horizon problem and seeds the initial density fluctuations for large-scale structure formation~\cite{Brout:1977ix,Sato:1980yn,Guth:1980zm,Linde:1981mu,Starobinsky:1982ee}, and is considered to be driven by a neutral scalar field~---~stable on cosmological scales~---~can also be a natural dark matter candidate. Several such possibilities have been considered in the literature so far ~\cite{Liddle:2006qz,Cardenas:2007xh,Panotopoulos:2007ri,Liddle:2008bm,Bose:2009kc,Lerner:2009xg,DeSantiago:2011qb,Khoze:2013uia,Mukaida:2014kpa,Fairbairn:2014zta,Bastero-Gil:2015lga,Kahlhoefer:2015jma,Tenkanen:2016twd,Daido:2017wwb,Choubey:2017hsq,Daido:2017tbr,Hooper:2018buz,Borah:2018rca,Manso:2018cba,Rosa:2018iff,Almeida:2018oid}. Generation of dark matter in the primordial Universe can be due to the incomplete decay of the inflaton (the remnant inflaton as the DM) or from the decaying inflaton~---~producing DM along with other SM degrees of freedom.  The fact that only one scalar field can minimally explain both inflation (explaining the temperature fluctuations as seen in the cosmic microwave background radiation (CMB)~\cite{Aghanim:2018eyx}) and make up the currently observed DM abundance in the Universe, is needless to say, an economical idea worth our attention.
\par
However, such a nonthermal production of dark matter is impossible to detect\footnote{Such non-thermal particle production was proposed to be tested via dark radiation or $N_{\rm eff}$ measurements during Big Bang Nucleosynthesis (BBN) \& CMB era, in context to neutrino anomalies \cite{Paul:2018njm}. } in any conventional astrophysical or laboratory-based dark matter searches, as they do not communicate with the visible sector (SM) and are typically considered to be beyond the paradigm of testability via any other interactions. Thankfully, the gravitational waves generated during the nonlinear preheating phase after inflation can provide an alternative channel for DM searches.
\par
The initial phase of inflationary (p)-reheating (see, e.g., Refs.~\cite{Kofman:1994rk,Kofman:1997yn,Mukaida:2013xxa}) is marked by resonant growth of the inflaton fluctuations and other fields that are coupled to the inflaton. The preheating soon gives rise to large inhomogeneties in the field fluctuations. These inhomogeneities act as a source of classical stochastic background gravitational waves~(SGWB) \cite{Khlebnikov:1997di,Easther:2006gt}. The production of such SGWB signals has been studied in literature over the years in various contexts such as single-field chaotic inflation~\cite{Easther:2006gt,Easther:2006vd, Easther:2007vj,Dufaux:2007pt}, preheating after hybrid inflation~\cite{Garcia-Bellido:2007nns,Garcia-Bellido:2007fiu,Dufaux:2008dn,Ringwald:2022xif}, including production of gauge fields during preheating~\cite{Dufaux:2010cf,Adshead:2018doq,Adshead:2019lbr}, and for GWs production from oscillons from inflationary potentials that \textit{`opens up'} away from the minimum~\cite{Antusch:2016con,Amin:2018xfe,Lozanov:2019ylm}.
\par
Inspired by the considerations mentioned above, this study addresses the issue of inflaton as dark matter, with a focus on the possibility of detection of the GW signals generated during the preheating phase. Our findings reveal that, with specific parameter choices, these GWs produced during preheating can be observable through GW detectors such as DECIGO, BBO, CE, and ET. Additionally, when exploring the inflaton as a candidate for dark matter, it places constraints on the parameter space of the inflaton coupling to fields. The significance of these alternative probes cannot be overstated, especially in contrast to traditional dark matter detectors, as they enable the exploration of non-thermal dark matter production mechanisms that involve limited or no interaction between the dark and visible sectors.
\par
The paper is organized as follows: 
In Section \ref{sec:secII}, we outline the basic prerequisites in the scalar field potential to ensure that the frequencies of the resulting GWs signals fall within the detectable spectrum. 
The challenges and prospects for achieving such potentials, which also satisfy the CMB observations on larger scales, are discussed in Appendix \ref{app:appA}.
We study the generation of classical sub-Hubble gravity waves from large time-dependent inhomogeneities due to (p)-reheating in Sec \ref{sec:secIV}. 
In Section \ref{sec:secV}, we present the specifics regarding the relic inflaton's role as dark matter, and we describe how to reheat the universe, keeping a stable inflaton relic while also making way for the impending radiation-dominated universe in Sec~\ref{sec:secVI}. 
Finally, we show the predictions of the GW spectrum in current and future detectors in Sec~\ref{sec:secVII}. 
This culminates in final discussions and conclusions in the concluding section.
We have used $\Mp=1/\sqrt{8\pi G}=2.435\times 10^{18}$GeV to denote the reduced Planck mass.
\section{\label{sec:secII}The scalar field potential during reheating}
Our preheating analysis will mostly be agnostic about the details of inflationary models during the inflationary phase. Our only requirement is that the field $\Phi$ responsible for generating the gravitational waves behaves as the archetypal $m_{\Phi}^2\Phi^2$ during the reheating phase. Most of the current observationally favored inflationary models such as the original Starobinsky model~\cite{Starobinsky:1982ee} in the Einstein frame or its recent generalization as the $\alpha$-attractor E-model~\cite{Carrasco:2015rva} or the T-model~\cite{Kallosh:2013yoa} with the exponent $n=2$ behaves as quadratic chaotic inflationary potentials. Other typical cases which can give rise to such behavior are:
\begin{enumerate}[leftmargin=*]
    \item[1] \textit{Hybrid inflationary models:} In this scenario, the inflaton responsible for generating the primordial perturbations can be of any shape~\cite{Linde:1993cn,Lyth:1998xn}. The waterfall field $\Phi$ dictates the end of inflation and is relevant for (p)-reheating. So, in such a scenario, we can identify the waterfall field $\phi$ and its inhomogeneities which give rise to the gravitational waves.
    \item[2] \textit{Curvaton models:} For such models, the field $\Phi$ is inflaton while the curvaton field seeds the primordial perturbations~\cite{Mazumdar:2010sa}.
\end{enumerate}
These theoretically well-motivated scenarios can alleviate the CMB constraints on the inflaton sector when looking for gravitational wave signatures from the primordial universe within the detectable range of various GWs detectors that are otherwise not accessible to CMB-based observations. 
The typical frequency of the observed GWs generated during preheating with a canonical inflationary model falls within ultrahigh-frequency (gigahertz) bands. 
However, as pointed out in~\cite{Easther:2006vd}, we can shift the frequencies towards the detectable ranges if we consider the effective mass $m_{\Phi}$ of the scalar field $\Phi$ as a free parameter.
In this regard, it is worth mentioning that the production of GWs driven by a system of coherent oscillating scalars taking the effective mass as a free parameter is also studied in the context of axions in~\cite{Kitajima:2018zco}. However, since we are considering that the scalar field is also responsible for the inflationary universe, we have more constraints on model building.
In the vanilla $m_{\Phi}^2\Phi^2$ model or other single field models such as $\alpha$-attractor models (where the effective mass around the minimum is determined by the amplitude of the plateau $\Lambda$ and the width $M$ of the potential), $m_{\Phi}$ is constrained from power spectrum normalization. 
For the remainder of the paper, we will use the $m_{\Phi}^2\Phi^2$ simplified model to exemplify the characteristics that emerge when treating the inflaton as dark matter and exploring the related GW signals from preheating. The details of preheating with the simple $m_{\Phi}\Phi^2$ model will follow in the next section. We will outline the challenges in model building in Appendix~\ref{app:appA}.

 \section{\label{sec:secIV}Gravitational Waves Generated During Preheating}
The preheating phase is generally episodic; the growth of the quantum fluctuations of the fields in the first phase is mainly due to the resonance phenomenon, while most of the system's energy is still in the homogeneous inflaton condensate, and the backreaction of the produced fields is not significant. The resonance, however, soon promptly populates all the modes of the quanta, and back-reaction becomes relevant. The second stage is where we observe the effects of back-reaction: the highly populated modes behave like matter waves and start colliding in a phase dominated by processes of nonlinear dynamics. These scatterings of waves happen in a highly relativistic and asymmetric manner. These processes generate a SGWB. Unlike the inflationary gravitational waves (tensor perturbations) generated from the quantum vacuum and get stretched to classical stochastic waves, GW production during preheating is a consequence of the scattering of the classical inhomogeneities, thus providing us with a complementary channel to test the physics of the early Universe. To fully capture the nonlinear dynamics of preheating and treat the backreaction carefully, we must study all the phases in full numerical simulation. Among several publicly available numerical codes to study preheating dynamics with $3$+$1$ dimensional lattice simulation~\cite{Felder:2000hq,Frolov:2008hy,Easther:2010qz,Huang:2011gf,Figueroa:2021yhd} we work with \textsc{PSpectRe}~\cite{Easther:2010qz} while taking the initial inhomogeneous seeds of the scalar fields to be generated similarly to \textsc{Defrost}~\cite{Frolov:2008hy}. Being a pseudospectral code, \textsc{PSpectRe} is free of differencing noise arising due to approximating the derivatives. In our simulation, the scalar field evolution is computed using a second-order-in-time symplectic integration scheme with a fixed time step as default in \textsc{PSpectRe}. We followed \cite{Easther:2007vj,Zhou:2013tsa} to compute the transverse-traceless part of the first-order metric perturbation. In synchronous gauge, the metric is written as:
\begin{equation}
    ds^2 = -dt^2 + a^2(t)\left(\delta_{ij} + h_{ij}\right)dx^idx^j,
\end{equation}
with $\partial_{ij} = h_{ij} = 0$. The equation of the motion of the metric fluctuations are:
\begin{equation}
    \ddot{h}_{ij} + 3H\dot{h}_{ij} - \frac{1}{a^2}\nabla^2 h_{ij} = \frac{2}{\Mp^2a^2}\Pi_{ij}^{TT},
\end{equation}
where $\Pi_{ij}^{TT}=\left[\partial_i\phi_a\partial_j\phi_a\right]^{TT}$ $(\text{with}~\phi_a=\{\phi,\chi\})$. 
\par
We then evolve the Fourier modes of the metric perturbations (and their sources in the stress-energy tensor) in Fourier space using a fourth-order-in-time Runge-Kutta scheme. The background evolution is self-consistently computed from the scalar field, only neglecting the backreaction of the metric fluctuations on the background. The GW energy density is given by
\begin{equation}
    \rho_{\rm GW}(t) = \frac{\Mp^2}{4}\langle\dot{h}_{ij}(\textbf{x},t)\dot{h}_{ij}(\textbf{x},t)\rangle_{\mathcal{V}},
\end{equation}
where $\langle\cdots\rangle_{\mathcal{V}}$ denotes a spatial average over $\mathcal{V}$.
Finally, the spectrum of the energy density of GWs (per logarithmic momentum interval) observable today is~\cite{Easther:2006gt,Easther:2006vd,Easther:2007vj,Dufaux:2007pt}
\begin{align}
\nonumber
\Omega_{\rm GW,0}h^2 &= \frac{h^2}{\rho_{\rm crit}}\frac{d\rho_{\rm GW}}{d\ln k}\Bigg|_{t=t_0} = \frac{h^2}{\rho_{\rm crit}}\frac{d\rho_{\rm GW}}{d\ln k}\Bigg|_{t=t_e}\frac{a_e^4\rho_{e}}{a_0^4\rho_{\rm crit,0}}\\
&= \Omega_{\rm rad, 0}h^2\Omega_{\mathrm{GW},e}\left(\frac{a_e}{a_{\ast}}\right)^{1-3w}\left(\frac{g_{\ast}}{g_0}\right)^{-1/3},
\end{align}
where, respectively, the subscript $0$ indicates quantities evaluated today, $e$ at the end of simulation and $\ast$ at the end of reheating. The equation of state~(EoS) of the universe between the end of inflation to the end of reheating is denoted by $w$. The critical density of the Universe is $\rho_{\rm crit} = 3\Mp^2H^2,$. Although the EOS at the end of reheating should ideally be radiationlike. However, this may not be the case for inflaton with a massive component as considered here but may reach values very close to $1/3$~\cite{Podolsky:2005bw,Maity:2018qhi,Saha:2020bis,Antusch:2020iyq}. Usually, perturbative reheating is considered in such cases to drain the inflaton energy density and bring the radiation-dominated universe. However, to identify the inflaton as a stable dark matter candidate, we require some mechanism to initiate the radiation-dominated universe without depleting the inflaton energy density. The details of such reheating mechanisms are described in Sec~\ref{sec:secVI}. Hence, for our calculation, we will take $w=1/3$. As a consequence, the redshifting of a {\it physical} wave vector, $k$, induced by preheating after inflation with corresponding energy scale $H_e$~(evaluated at $a_e$), corresponds to an observed frequency today as
\begin{equation}
    f = 1.32\times10^{10}\frac{k}{\sqrt{M_{\rm Pl}H_e}}
\end{equation}
\par
Now, for the simulation we will work with $V(\Phi, |H|) = \frac{1}{2}m_{\Phi}^2\Phi^2 + \frac{1}{2}g^2\Phi^2|H|^2$ scenario where $|H|$ is produced due to inflaton decay. The dimensionless coupling $g$ is generally taken to be small to avoid large radiative corrections during inflation. After the end of inflation, when the inflaton oscillates about the minimum of the potential, we can identify the effective mass parameters for any model that behaves as $V(\Phi)\propto\Phi^2$ around the minimum. As mentioned earlier, we will take $m$ as a free parameter; additionally, to have a sufficient amplitude of GWs, we also need to take the initial amplitude of the inflaton to be large enough ($\sim \Mp$). This makes the model building a bit challenging in this case. For instance, in $\alpha$-attractor models, to decrease the effective mass, we will also need to decrease the value of the $\alpha$ parameter. It, however, will decrease the initial amplitude of the inflaton field ($\Phi_0$) at the start of the simulation. We mention some directions on the inflationary model-building aspects in the Appendix. The essential requirement for unifying the inflaton and dark matter sector is that of incomplete decay of the inflation, which is characteristic of preheating with massive inflaton.
\section{\label{sec:secV}Inflaton as the Dark Matter}
We describe the post-inflationary dynamics of the inflaton field in this section: the end of inflation is marked by the oscillation of the inflaton field around the minima of the potential minimum, and during this time, due to oscillations of the inflaton field, the energy density of the inflaton get transferred to other fields that are coupled to the inflaton. Due to the inherent $Z_2$ parity, under which the inflaton transforms as odd, the inflaton cannot decay to any other particles, so perturbative reheating is not possible, following the preheating particle production \cite{Kofman:1994rk} which in this case happens via, the $g^2 \Phi^2 |H|^2$ coupling, we expect the inflaton to violently produce $\Phi$ particles which are also known as the parametric resonance, due to Bose-Einstein scalar field statistics. When the inflaton amplitude becomes smaller than $m_{\Phi}/g$, the preheating stops \cite{Kofman:1994rk,Kofman:1997yn}.
\par
The oscillating mode of the stable inflaton stops transferring the energy density into any other sector at the end of the preheating era and thus may act as the dark matter, with the ratio of the average energy density of this oscillation mode to the number density of photons being given by \cite{Liddle:2008bm,Okada:2010jd}
\begin{align}
\xi_{{\rm dm},0} \equiv \frac{\rho_{\Phi,0}}{n_{\gamma,0}}\simeq 0.04 \left( \frac{m_\Phi}{\Mp}\right)^{1/2}
  \left( \frac{\Phi_*}{\Mp}\right)^2 \Mp,
  \label{eq:dm_to_photon}
\end{align}
where $\Phi_*$ is the oscillation amplitude at time $t_*$ when $m_\Phi=H_*$, the Hubble parameter. We have also used the typical values of the total number of relativistic particle degrees of freedom $g_{\ast}\simeq 100$ and preset entropic degrees of freedom $g_{S,0}=3.9$.
The averaged inflaton energy density for $t>t_{\ast}$ scales as $\rho_{\Phi} = \frac{1}{2}m^2\Phi_{\ast}^2a_{\ast}^3/a^3$. 
With $\Phi_* = m_{\Phi}/g $, we find $Y_{{\rm dm},0} \simeq 7.5 \times 10^{-37} \Mp$ for $m_{\Phi}=1$ TeV and $g^2=0.1$. 
To satisfy the observed relic density of the universe, $\xi_{\mathrm{dm},0} \simeq 5.5 \times 10^{-27} \Mp$, and therefore we conclude that the oscillating mode has a negligible contribution to the energy density of the present Universe\footnote{It is worth recalling that successful $m_{\Phi}^2 \Phi^2$ chaotic inflaton requires the inflaton mass $m_{\Phi} \simeq 10^{13}$ GeV, and so $g \simeq 10^7$ is needed to realize the observed value of $\Omega_{\mathrm{dm},0}$ \cite{Liddle:2008bm}. However, in our case, as the inflaton mass around the minima is a free parameter, we can take $m_{\Phi} \ll 10^{13}$ GeV.}.
\par
One crucial issue we must address is how to reheat the universe while keeping the remnant inflaton relic as a stable dark matter. 
In the upcoming section, we will address the matter of perturbative reheating specifically tailored for scenarios involving the inflaton as dark matter.

\section{\label{sec:secVI}Incomplete decay of Inflaton and Reheating}
Preheating after inflation with $m_{\Phi}^2\Phi^2$ type potential around the minima with four-legged interactions, as we have considered in this work, is known to produce an incomplete decay of inflaton~\cite{Podolsky:2005bw,Maity:2018qhi,Saha:2020bis,Antusch:2020iyq}. Although including a three-legged decay term can lead to a complete decay~\cite{Dufaux:2006ee}, we bar such interactions by presuming that the $Z_2$ is (almost) exact symmetry for the inflaton. Thus, under these circumstances, we cannot expect the radiation-dominated era to commence; rather, a phase of a matter-dominated era reemerges after a few $e$-folds~\cite{Podolsky:2005bw,Maity:2018qhi,Saha:2020bis,Antusch:2020iyq}.
In such cases, once preheating becomes inefficient (once the inflaton amplitude drops below $m_{\Phi}/g$ or $\Phi_0/\sqrt{q}$), the perturbative reheating with explicit inflaton decay terms are invoked for a complete decay of the inflaton and to bring in the radiation-dominated universe. But if the inflaton is to remain as the stable dark matter relic, we must forbid such perturbative decay. Several proposals are made in the literature in this direction:
\begin{itemize}
    \item In~\cite{Cardenas:2007xh,Panotopoulos:2007ri}, the authors utilized the idea of plasma mass effects~\cite{Kolb:2003ke} that kinematically forbids the inflaton decay in some part of the reheating phase. However, inflaton can decay in such cases once it is subdominant with respect to the radiation fluid. Consequently, such scenarios will not be helpful for the case of a unified description of inflaton as dark matter.
    \item Authors in~\cite{Liddle:2008bm} propounded the idea of triple-unification by proposing a thermal inflation period following preheating. This phase of second inflation reduces the amount of inflaton/dark matter to the observed level. However, such a phase inflation of thermal inflation will not be suitable when considering the production of GWs from preheating.
    \item In~\cite{Bastero-Gil:2015lga}, the authors noticed that as the mass of the inflaton decay products is field dependent, employing it is possible to construct models where inflaton decay only during the initial stage of coherent oscillation while it is kinematically forbidden at late times. By employing a suitable discrete symmetry that facilitates such decay away from minima of the potential, they can show that the scenario can bring in radiation-dominated era while the inflaton particles thermalize in the process and eventually decouple and freeze out, yielding correct DM relic abundance. This scenario has been further considered for identifying the inflaton field as dark energy at late times in the context of warm inflation~\cite{Rosa:2018iff}.
\end{itemize}
Below, we will outline, following~\cite{Bastero-Gil:2015lga}, the appropriate reheating scenario to bring in the radiation-dominated universe with a stable inflaton acting as the dark matter relic.  Here, we consider the inflaton is also coupled to fermionic fields $\Psi_{+}$ and $\Psi_{-}$ via the standard Yukawa interactions, and they satisfy the following discrete symmetry $C_2 \subset \mathbb{Z}_2 \times S_2$ such that when the inflaton transform as $\Phi\leftrightarrow -\Phi$ the fermions are simultaneously transformed under the permutation symmetry as $\Psi_{+} \leftrightarrow \Psi_{-}$. The resulting Lagrangian density is
\begin{align}
\nonumber
    \mathcal{L} =& \frac{1}{2}\partial_{\mu}\Phi\partial^{\mu}\Phi - V(|\Phi|) - \frac{1}{2}\Phi^2|H|^2\\
    \nonumber
    &+ \bar{\Psi}_{+}\left(i\gamma^{\mu}\partial_{\mu} - m_f\right)\Psi_{+} + \bar{\Psi}_{-}\left(i\gamma^{\mu}\partial_{\mu} - m_f\right)\Psi_{-}\\
    &- h\Phi\bar{\Psi}_{+}\Psi_{+} + h\Phi\bar{\Psi}_{-}\Psi_{-} 
\end{align}
Now, because of the above discrete symmetry, the two fermions have the opposite Yukawa coupling to the inflation while having the same tree-level mass $m_f$. The discrete symmetry is preserved at the minimum of the potential at $\Phi=0$; consequently, perturbative decay to the fermionic channel is only allowed. When $m_f > m_{\Phi}/2$, these decays are kinematically blocked, and the inflaton is completely stable, which accounts for the present DM abundance. 
\par
Away from minimum, the discrete symmetry is broken, causing mass splitting of the fermions $m_{\pm} = |m_f \pm h\Phi|$. 
Perturbative decay is allowed when field values satisfy $|\Phi| \gtrsim m_{f}/h$. 
We also note that due to Fermi blocking, we could neglect the resonance effects of the fermions during preheating. 
Additionally, the nature of the interaction makes it possible to neglect the fermionic interaction at the early stages of the preheating when $m_{\pm}\sim h|\Phi|\gg m_{\Phi}$, i.e., at the time of the production of GWs due to the parametric resonance via four-legged interactions.
\par
To sum up the reheating epoch, we have the following scenario for the successive phases of inflaton decay: (i) preheating commences after the end of inflation when $H\sim m_{\phi}$ with an initial amplitude of the field $\Phi_0$. Until the amplitude drops below $\Phi_0/\sqrt{q}$, preheating is efficient and serves as the primary channel for energy transfer from the inflaton to the decay products. (ii) When $m_{\Phi}/g \gtrsim |\Phi| \gtrsim m_f/h$, we will have perturbative inflaton decay via fermionic channels. (iii) Perturbative decay is prohibited, thanks to the discrete symmetry mentioned above, when inflaton amplitude drops below $m_f/h$, allowing a stable relic inflaton component.
\par
\begin{figure}[ht]
    \centering
    \includegraphics[scale=0.35]{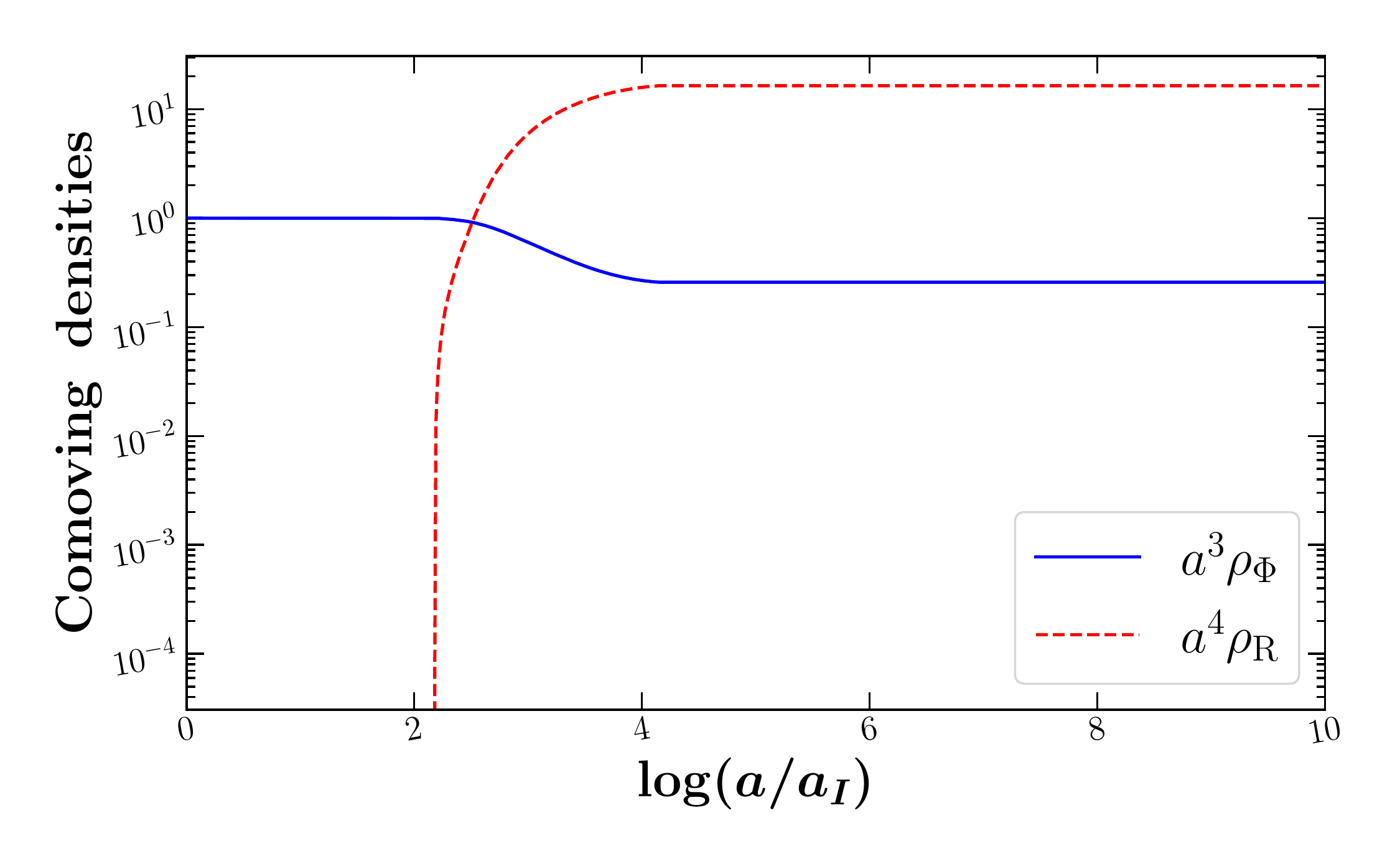}
    \caption{We plot the comoving density of the inflaton (blue) and radiation (red-dashed) during the perturbative reheating
process after preheating. The radiation-dominated universe commences with a stable inflaton relic working as the dark matter
candidate.}
    \label{fig:pert_decay}
\end{figure}

Now, the total decay width of the processes is $\Gamma_{\Phi} = \Gamma_{+} + \Gamma_{-}$ with the partial decay widths of the fermionic decay channels given by
\begin{equation}
    \Gamma_{\pm} = \frac{h^2}{8\pi}m_{\Phi}\left(1 - \frac{4m_{\pm}^2}{m_{\Phi}^2}\right),
\end{equation}
Then we solve the following Boltzmann equations for the perturbative inflaton decay
\begin{align}
\label{eq:pert_inf}
    \dot{\rho}_{\Phi} + 3H(\rho_{\Phi} + p_{\Phi}) &= -\Gamma_{\Phi}\dot{\Phi}^2,\\
    \label{eq:pert_rad}
    \dot{\rho}_{R} + 4H\rho_{R} &= \Gamma_{\Phi}\dot{\Phi}^2,
\end{align}
with the expansion given by the following Friedmann equation
\begin{align}
\nonumber
        3\Mp^2H^2 &= \rho_{\Phi} + \rho_{R},\\
                  &= \frac{1}{2}\dot{\Phi}^2 + V(\Phi) + \rho_{R}.
\end{align}
In Fig. (\ref{fig:pert_decay}), We show the evolution of the densities of inflaton $\rho_{\Phi}$ and the radiation bath $\rho_{R}$ at temperature $T$ [$\rho_{R} = (\pi^2/30)g_{\ast}T^4$ with $g_{\ast}$ being the effective number of relativistic degrees of freedom forming radiation bath]
The reheating temperature is given by
\begin{align}
\nonumber
    T_{R} &\simeq \left(\frac{15\sqrt{6}}{16\pi^4}\right)^{1/4}g_{\ast}^{-1/4}h^{1/4}\left(\frac{m_{\Phi}}{\Mp}\right)^{3/4}\Mp,\\
    &\simeq 2.7\times10^{6}g_{\ast}^{-1/4}h^{1/4}\left(\frac{m_{\Phi}}{1\mathrm{TeV}}\right)^{3/4}\mathrm{GeV}.
    \label{eq:tre}
\end{align}
Thus, the typical amplitude when perturbative inflaton decay is forbidden can be tuned by choosing the fermion mass and the Yukawa interaction. 
For a typical value of $m_{\Phi}$ for detectable GWs, we see from~(\ref{eq:tre}) that the reheating temperature will respect the BBN bound~\cite{Kawasaki:1999na,Kawasaki:2000en,Steigman:2007xt}. With an understanding of how to reheat the universe, keeping a stable inflaton relic, we will describe the details of our numerical simulation in the following section.
 \section{\label{sec:secVII}Numerical Analysis}
 \label{sec:numerics}
\begin{figure*}[t]
     \centering
     \includegraphics[scale=0.8]{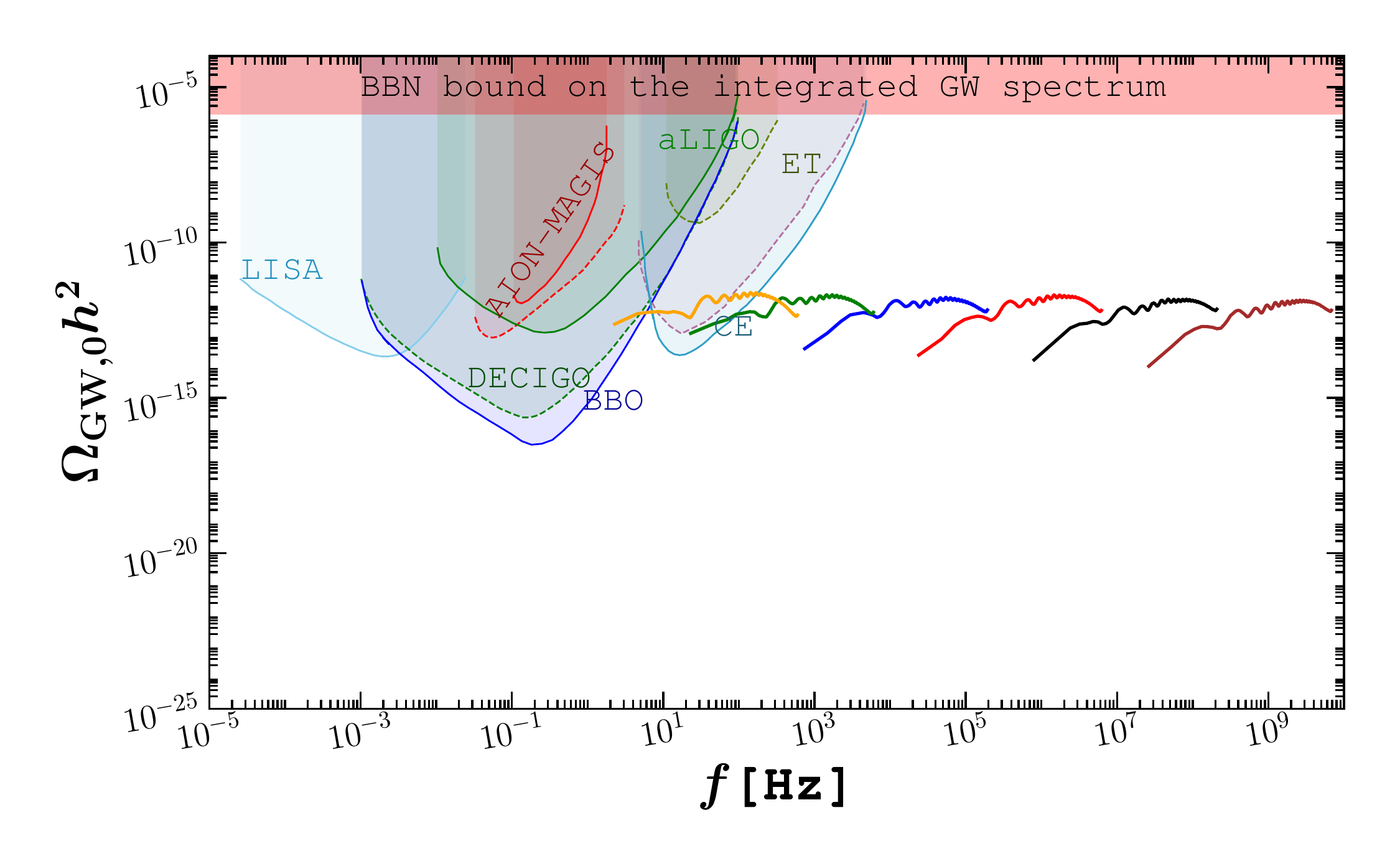}
     \caption{\it Spectrum of gravitational waves generated during preheating projected on the detection sensitivity of various experiments. With take the dimensionless resonance parameter as $(g^2\Phi_0^2)/m_{\Phi}^2=10^4$ and start our simulation when $\Phi_0=1~\Mp$ for all the cases. We change the energy scale, keeping $m$ as a free parameter with this initial setup. We start with $m_{\Phi}=5\times10^{-6}~\Mp$~(red in the extreme right) which corresponds to an energy scale of $(4.5\times 10^{15}~{\rm GeV})^4$ to $m_{\Phi}=5\times10^{-20}~\Mp$~(black in the extreme left) which corresponds to a energy scale of $(4.5\times 10^{8}~{\rm GeV})^4$. This final value of $m_{\Phi}=120~{\rm MeV}$ is chosen such that it is well above the bound on the minimum reheating temperature of $10~{\rm MeV}$ from BBN predictions. The intermediate curves are obtained choosing the values of $m_{\Phi}$ to be $5\times10^{-9}~\Mp$~(blue),~$5\times10^{-12}~\Mp$~(green),~$5\times10^{-15}~\Mp$~(yellow) and $5\times10^{-18}~\Mp$~(brown). The gravitational waves generated during low-scale inflation fall within the sensitivity range of detectors such as DECIGO, BBO, LIGO, etc.}
     \label{fig:fig1}
 \end{figure*}

\begin{figure*}[t]
     \centering
     \includegraphics[scale=0.75]{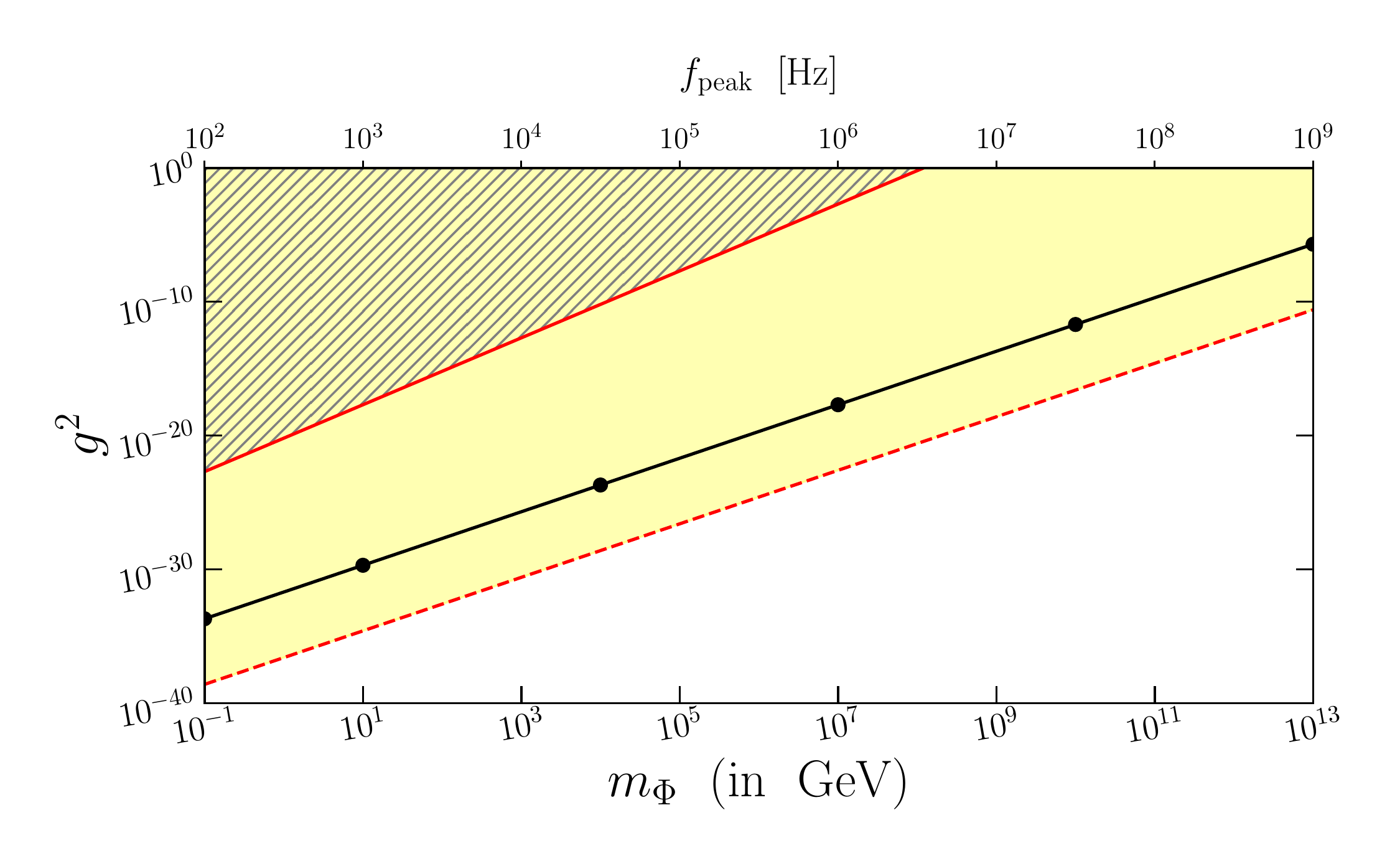}
     \caption{\it Plot shows the parameter space in the $m_{\Phi}$-$g^2$ plane, which will satisfy the observed dark matter to photon ratio (redline). For values of the parameters, values from hatched are allowed but will produce less $Y_{dm,0}$ in that case; the inflaton will not account for the full DM. The parameters within the yellow-shaded region correspond to $q>1$, which naively tells us the region where broad resonance is possible. However, in practice, for broad resonance, one must have $q\gg1$. We took $q=8\times10^{4}$. The black dots correspond to the values used in our simulation. The frequency in the upper axis corresponds to the peak frequency of the GWs signal using the fitted relation in~(\ref{eq:fitfreq}).}
     \label{fig:fig2}
 \end{figure*}

 We performed our simulations with $512^3$ lattice points. We kept the dimensionless resonance parameter $q( =g^2\Phi_0^2/m_{\Phi}^2)$ fixed at $q = 8\times10^4$ for all the simulations. Where the initial values of inflaton oscillation $\Phi_0$ are taken to be $1~\Mp$. We show the produced GWs spectra in Fig~(\ref{fig:fig1}). As mentioned earlier, we are interested in the scenarios where the CMB observables do not constrain the quadratic potential during preheating; consequently, we choose the inflaton mass $m_{\Phi}$ as a free parameter. We have taken its value for each run as $\{5\times10^{-6}~\Mp,5\times10^{-9}~\Mp,5\times10^{-12}~\Mp,5\times10^{-15}~\Mp,5\times10^{-18}~\Mp,5\times10^{-20}~\Mp\}$. With these values, the ratio  of the average energy density of the oscillation modes to the number density of photons defined in (\ref{eq:dm_to_photon}), are found to be $\{8.8\times10^{-9} \Mp, 2.8\times10^{-10} \Mp, 8.8\times10^{-12} \Mp, 2.8\times10^{-13} \Mp, 
 8.8\times10^{-15} \Mp, 8.8\times10^{-16} \Mp \}$. Now to satisfy the measured value of the current dark matter per photon $\xi_{\rm dm, 0} = 1.1\times10^{-27}\Mp$, we require the value of coupling $g$ to have values $\{1.4\times10^6, 2.5\times10^2, 4.5\times10^{-2}, 8\times10^{-6}, 1.4\times10^{-9}, 
 4.5\times10^{-12}\}$. 
 We have plotted the results in (\ref{fig:fig2}). The solid red line in the $m_{\Phi}$-$g^2$ parameter space satisfies the observed dark matter to photon ratio. Notice that for the cases considered here, when we have inflaton mass below $3\times10^{-11}\Mp$ or $7.34\times10^4$ GeV, the couplings are well within the perturbation limit. However, we have restricted our simulations for those values of $q$, providing us with sufficient resonance without any numerical instabilities. 
 Simulations for all $g^2$ values along the red line in Fig.(\ref{fig:fig2}) will result in a large resonance parameter, rendering the lattice simulation afflicted by numerical instabilities. 
 Nevertheless, provided we have sufficient resonance; the amplitude of the GW spectrum only depends on the strength of the resonance parameter while the frequency range depends on the initial energy scale. 
 Consequently, for the same initial energy, changing the coupling will produce a `tower' of GW spectrum within the same frequency range~\cite{Hyde:2015gwa,Cai:2021gju,Figueroa:2022iho}. 
 Thus, we can say that those values of the coupling $g^2$ along the red line larger than the simulations performed here will also ensure parametric resonance. Consequently, the corresponding GW signals should also be detectable with various detectors (see Fig. \ref{fig:fig1}). 
 This implies that the coherently oscillating scalar field during (p)-reheating that behaves as $m_{\Phi}^2\Phi^2$ type can source a promising candidate for the dark matter. 
Additionally, we would like to comment that if we consider the production of GWs from coherent scalar while relaxing the scalar field as inflaton, we can have detectable GW signals across a much wider range of frequency bands~\cite{Cui:2023fbg}. 
 For instance, fitting the peak frequency of the red-shifted GWs signal with the energy scale and hence the effective mass, we find the following relation:
    \begin{equation}\label{eq:fitfreq}
    f_{\mathrm{peak}}  = 3.16\times 10^2\left(\frac{m_{\Phi}}{\mathrm{GeV}}\right)^{1/2}\mathrm{Hz}
    \end{equation}
    Hence, depending on the scalar field mass, we can move to the low-frequency GWs detector bands. 
    All these scenarios could be interesting to explore in light of recently detected GWs background by the North American Nanohertz Observatory for Gravitational Waves (NANOGrav)~\cite{NANOGrav:2023gor}.
 A comprehensive analysis of the amplitude of GW spectrum involving various interactions between the inflaton and the Higgs field, along with the effects of various initial conditions and a broad range of parameters of interest, will be examined in a future publication (also see Refs.~\cite{Hyde:2015gwa,Sang:2019ndv,Cai:2021gju,Ashoorioon:2022raz,Figueroa:2022iho} for similar works along this direction).
\medskip
 
\section{\label{sec:disc}Discussion and Conclusion}
In a nutshell, we studied the inflaton behaving as the dark matter of the Universe. We summarize our main findings below:

\begin{itemize}
    \item We showed in a (p)-reheating scenario, the inflaton starts to oscillate, and its zeroth mode energy remnant behaves as nonthermal dark matter as our observable current Universe satisfying the relic density (cf. Eq.~\ref{eq:dm_to_photon}). We investigated the impact of resonance parameters on dark matter production via lattice simulation in the oscillating regimes for such an oscillating scenario.
    \item In such a nonthermal origin of dark matter without any interaction with the visible or the SM sector, it is impossible to have laboratory tests for the DM candidate. We showed that the generated gravitational waves could be a viable detection channel during such an oscillatory period with its typical spectrum, as shown in Fig \ref{fig:fig1}.
        \item In Fig \ref{fig:fig2}, we showed the allowed parameter space of the coupling $g^2$ and the effective mass of the coherent oscillating inflaton $m_{\Phi}$, which satisfy the observed DM mass to photon ratio. If we have sufficient resonance, the amplitude of the produced GWs is independent of the energy scale of the inflation (cf Fig (\ref{fig:fig1})). Thus, we found that for $m_{\Phi}\lesssim 10^5$GeV, the generated stochastic signals of GWs will be of sufficient amplitude to be detectable in various planned and proposed detectors while also explaining the origin of dark matter as the remnant inflaton component.
    \item As depicted in Fig 2, for the higher mass range ($m_{\Phi}\gtrsim 10^8$GeV), the coupling required to satisfy the observed dark matter relic is excluded by perturbative limit. Moreover, the allowed coupling range will overproduce dark matter and, hence, not allowed when considering inflaton as dark matter.
    For the lower end, although the couplings required to satisfy the dark matter relic are within the perturbative limit, the resulting resonance parameters became so large that the simulations became numerically unstable. However, in principle, the GWs produced in such parameter regions should also be detectable. Thus, the lower-mass ranges~$m_{\Phi}\lesssim 10^8$~GeV provide an interesting parameter space for the DM probe using GWs detectors.
    \item We can probe mass scales ($\geq$ TeV) otherwise not accessible to colliders or any other laboratory physics. There is huge interest in the community to find the detection principles of high-frequency gravitational gaves~\cite{Aggarwal:2020olq,Berlin:2021txa,Domcke:2022rgu}. With more advanced detectors, although we may be able to probe high-scale inflationary scenarios, the paradigm of inflation as the dark matter may be in tension within the scope of simple four-legged interaction as considered here.
\end{itemize}

We have thus explored a novel alternative to the thermal WIMP dark matter scenario in context to the non-thermal production of dark matter in the early Universe. The inflaton itself starts to behave as dark matter in late times, with the energy density stored during the oscillations of the (p)-reheating era. Due to no other or negligible interactions being present between the inflaton sector (dark sector) and the visible sector, we showed our prospects of detecting such non-thermal dark matter candidates within the measurements of SBGW in current and future GW detectors.
We envisage our results will be the commencement of huge activity in this direction, whereas an alternative to hunt for various non-thermal dark matter production mechanisms which are otherwise not traceable in laboratory-based or astrophysical searches and, therefore, not in traditional DM detectors due to very feeble or no coupling between the visible and the dark sectors but in GW detectors in the near future, particularly when LIGO has already put bounds \cite{2009Natur.460..990A}, and LIGO-VIRGO has further searched for such SGWB \cite{KAGRA:2021mth} from primordial sources as described in \cite{Romero:2021kby}.

\medskip
\section*{Acknowledgments}
Authors thank Rouzbeh Allahverdi, Nicolas Bernal, Lucien Hertier, 
Tomasz Krajewski, Marek Lewicki, Anupam Mazumdar, Kyohei Mukaida, Myeonghun Park, and Yuko Urakawa for helpful comments. 
The study of P. S. is financially supported by the Seoul National University of Science and Technology. 
The simulations are performed in \texttt{Dark} cluster at Seoultech. 

\appendix
\section{\label{app:appA}Towards Inflationary model building for reheating with detectable GWs signal}
The effective mass in most canonical inflationary models is constrained by the amplitude of the scalar spectrum at the CMB scales. For instance, in the simple one-parameter chaotic $m^2\Phi^2$ potential, the value of the mass $m$ is fixed at $10^{-6}\Mp$. Looking beyond, the extra parameter models may provide us with some freedom. The currently observationally favored alpha-attractor models, for instance, the super gravity-inspired T-models having the inflationary potential $V(\Phi) = \Lambda^4\tanh^2(\phi/M)$ reduces to $V(\Phi) \propto (\Lambda^4/M^2)\Phi^2$ around the minima. Hence, the effective mass can be written as $m_{\rm eff}^2 = \Lambda^4/M^2$, where $\Lambda$ is the amplitude of the potential’s plateau and $M$ is a free mass scale describing the width of the potential. In this case, however, it turns out that the observed amplitude of the scalar perturbations imposes approximately $\Lambda^4 \propto M^2$, so we have $m_{\Phi} \approx 10^{-6}\Mp$ for all values of $M$~\cite{Antusch:2021aiw}. Consequently, the effective mass $m_{\Phi}$ is the same as the vanilla chaotic inflation~\footnote{We thank the referee for pointing this out to us.}. Thus, we must expand our search for suitable models in other directions. We list below some models where we can take the effective mass during preheating as a free parameter and describe the associated challenges.

\subsection{Two-field hybrid inflation}
To make the effective mass a free parameter, the authors in~\cite{Easther:2006vd} assumed that the full potential during inflation is given by:
\begin{equation}
    V = \frac{M^2 - \lambda\sigma^2}{4\lambda} + \frac{1}{2}m^2\Phi^2 + \frac{1}{2}h^2\Phi^2\sigma^2 + \frac{1}{2}g^2\Phi^2\chi^2.
    \label{eq:pot_East}
\end{equation}
During inflation, when $\Phi>M/h$, $\sigma$ stays in its false $\sigma=0$. Thus, the effective potential during inflation is
\begin{equation}
    V(\phi) = \frac{M^4}{4\lambda} + \frac{1}{2}m^2\Phi^2.
    \label{eq:pot_East_inf}
\end{equation}
The auxiliary field $\sigma$ becomes tachyonic when $\Phi=M/h$ and evolves towards its true vacuum at $M\sqrt{\lambda}$. Assuming $\sigma = \langle\sigma\rangle$, the potential reduces to
\begin{equation}
        V(\Phi) = \frac{1}{2}\left(m^2 + \frac{h^2M^2}{\lambda}\right)\Phi^2.
    \label{eq:pot_East_eff}
\end{equation}
Now, in the limit $m^2\ll h^2M^2/\lambda$, The effective potential is given by
\begin{equation}
    V(\Phi) = \frac{1}{2}\mu^2\Phi^2,
\end{equation}
where $\mu^2\equiv h^2M^2/\lambda$, and, consequently, the effective mass for the inflaton can be taken as a free parameter. However, if $m^2\ll h^2M^2/\lambda$ and during inflation when $\Phi>M/h$, the potential in~(\ref{eq:pot_East_eff}) will result in blue tilted scalar spectra~($n_s=1$). Although such blue-tilted spectra have implications for resolving the Hubble tension~\cite{Takahashi:2021bti,Jiang:2022uyg,Ye:2022efx}, such spectra are ruled out from observations.

\subsection{A model in the framework of $\alpha$-attractor}
A broad class of inflationary models defined in terms of a parameter $\alpha$ in the context of supergravity are the well-known $\alpha$-attractors. The aforementioned parameter determines the curvature and cutoff of these models. These models are known to predict some universal results of the inflationary observables, viz., the spectral index ($n_s$) and the tensor-to-scalar ratio ($r$). There are several inequivalent formalisms to introduce the cosmological $\alpha$-attractors which eventually leads to the following Einstein frame Lagrangian~\cite{Kallosh:2014rga}:
\begin{equation}
    \mathcal{L} = \sqrt{-g}\left[\frac{1}{2}R - \frac{\alpha(\partial\phi)}{\left(1-\frac{\phi^2}{6}\right)^2}-f^2\left(\frac{\phi}{\sqrt{6}}\right)\right],
    \label{eq:alpha_att0}
\end{equation}
which after rescaling of the field $\phi$
\begin{equation}
    \mathcal{L} = \sqrt{-g}\left[\frac{1}{2}R - \frac{(\partial\phi)}{\left(1-\frac{\phi^2}{6\alpha}\right)^2}-f^2\left(\frac{\phi}{\sqrt{6\alpha}}\right)\right],
    \label{eq:alpha_att}
\end{equation}
Finally, using the scalar field redefinition $\phi\sqrt{6\alpha} = \tanh(\Phi/\sqrt{6\alpha})$, we arrive at the following form of Lagrangian in the Einstein frame with a canonically normalized kinetic term:
\begin{align}
    \mathcal{L}_{\alpha} = \sqrt{-g}\Bigg[\frac{1}{2}R(g) - \frac{1}{2}\partial_{\mu}\Phi\partial^{\mu}\Phi - f^2\left(\tanh\frac{\Phi}{\sqrt{6\alpha}}\right)\Bigg].
\end{align}
\begin{equation}
    f(x) = f_0\left[\frac{x^n + A\Theta(x-x_{\rm Step})}{x_{\rm CMB}^n + A\Theta(x_{\rm CMB} - x_{\rm Step})}\right]
\end{equation}
which will have $n$ step-like transitions. We will only concentrate on the $n=1$ case for the rest of our analysis. 
\begin{figure}[!t]
    \centering
    \includegraphics[scale=0.5]{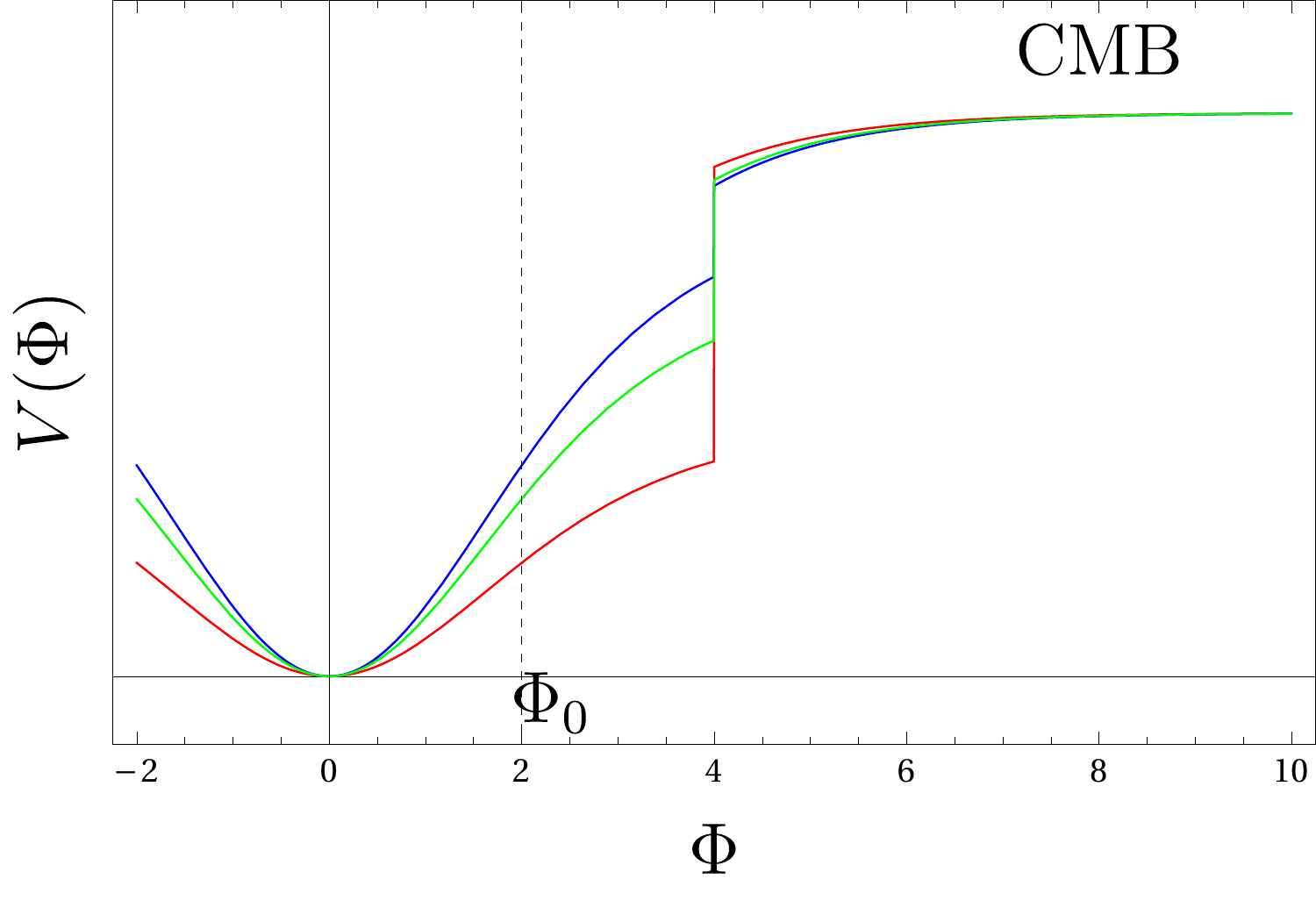}
    \caption{An illustration of the scalar potential for $\alpha=1$ and $A=0.5$~(red), $0.05$~(green) and $0.1$~(blue). The height of the potential at the position of the step determines the effective mass of the potential around the minima.}
    \label{fig:scalar_pot}
\end{figure}
We take $x_{\rm CMB} = \tanh\frac{\Phi_{\rm CMB}}{\sqrt{6\alpha}}$ that determines the normalization of the potential in the flat region to for matching the predictions of the model with the CMB observables. $x_{\rm Step} = \tanh\frac{\Phi_{\rm Step}}{\sqrt{6\alpha}}$ determines the appearance of the step.
The inflationary potential is found to be:
\begin{align}
\nonumber
    V(\Phi) &= f^2\left(\tanh\frac{\Phi}{\sqrt{6\alpha}}\right),\\
            &= V_0\left[\frac{\tanh\frac{\Phi}{\sqrt{6\alpha}} + A\Theta\left(\tanh\frac{\Phi}{\sqrt{6\alpha}}-\tanh\frac{\phi_{\rm Step}}{\sqrt{6\alpha}}\right)}{\tanh\frac{\Phi_{\rm CMB}}{\sqrt{6\alpha}} + A\Theta\left(\tanh\frac{\Phi_{\rm CMB}}{\sqrt{6\alpha}}-\tanh\frac{\phi_{\rm Step}}{\sqrt{6\alpha}}\right)}\right]^2
\end{align}
We have expressed all the dimensionful quantities in the units of $\Mp$ and $V_0=f_0^2$. In practice, we can use continuous step function, for instance, $\Theta(x-x_0) = (1+\tanh((x-x_0)/\Delta))/2$ for numerical stability when solving the equations for the background inflationary dynamics and perturbations.
The effective mass around the minima:
\begin{align}
    m_{\Phi}^2 = \frac{V_0}{6\left[A\Theta\left(\tanh\frac{\Phi_{\rm CMB}}{\sqrt{6\alpha}}-\tanh\frac{\Phi_{\rm Step}}{\sqrt{6\alpha}}\right) + \tanh\frac{\Phi_{\rm CMB}}{\sqrt{6\alpha}}\right]^2}
\end{align}
Now, being derivatives of the $\alpha$-attractor models, these models with steps can easily produce the necessary CMB observables. For instance, taking $\alpha=1$ and $A=0.05$ we can have $n_s=0.9630$ and $r_{0.05}=0.003$ while the amplitude of power spectrum fixes the value of $V_0$ to be $1.344\times10^{-10}$. This will make the effective mass around the minima $m_{\Phi}=4\times 10^{-6}\Mp$. Increasing $A$ will decrease the $m_{\Phi}$ but will require tuning the other parameters. In all cases, the only requirement for the preheating scenario to work is that the position of the step is at higher field values compared to the initial amplitude of the inflaton for simulation ($\Phi_0=1\Mp$). In the above case, we have taken $\Phi_{\rm CMB}=10$ and $\Phi_{\rm Step}=4$. Such models can also boost the scalar power spectrum, which has interesting applications for Primordial black holes and secondary gravitational waves~\cite{Kefala:2020xsx,Dalianis:2021iig}. It is important to mention that to make $m_{\Phi}$ as low as in the MeV range for the above model, we will need to take an extremely large height of the step. This may result in a kination phase after the end of inflation that can make the relation between the effective mass of inflaton and the observed GWs signal more convoluted. 
\par
At this point, we must emphasize that the models listed above are only suggestive ways of making the effective mass a free parameter and looking for a more realistic model is still an open issue. We adjourn the details of model building and parameter explorations for future work. 

\providecommand{\href}[2]{#2}\begingroup\raggedright\endgroup

\end{document}